\documentstyle[12pt]{article}
\def\lsim{\mathrel{\rlap{\lower4pt\hbox{\hskip1pt$\sim$}}
    \raise1pt\hbox{$<$}}}         
\def\gsim{\mathrel{\rlap{\lower4pt\hbox{\hskip1pt$\sim$}}
    \raise1pt\hbox{$>$}}}         

\def\overleftrightarrow#1{\vbox{\ialign{##\crcr
    $\leftrightarrow$\crcr
    \noalign{\kern 1pt\nointerlineskip}
    $\hfil\displaystyle{#1}\hfil$\crcr}}}

\begin{document}
\begin{center}
 
{\bf Addendum to the Weak Parity-Violating Pion-Nucleon Coupling}
\vspace{.5in}
 
E.M. Henley \\
 
{\em Department of Physics and Institute for Nuclear Theory, \\
Box 351560, University of Washington, Seattle, Washington 98195}
 
\vspace{2mm}
W-Y.P. Hwang \\
 
{\em Department of Physics, National Taiwan University, \\
Taipei, Taiwan 10764, \\
and \\
Center for Theoretical Physics, Laboratory for Nuclear Science \\
and Department of Physics, Massachusetts Institute of Technology, \\
Cambridge, Massachusetts 02139} 
 
\vspace{2mm}
L.S. Kisslinger \\
{\em Department of Physics, Carnegie-Mellon University, \\
Pittsburgh, Pennsylvania 15213}

\end{center} 
\vspace{.5in}

    We need to make a correction to our earlier work \cite{HHK}, namely
a critical error in the relative sign between the lowest-dimensional term
and the leading nonperturbative contribution. By correcting the sign error and
evaluating an additional contribution which we omitted earlier, we conclude
that our new prediction on the weak pion-nucleon coupling $f_{\pi NN}$
is in the vincinity of $3\times 10^{-7}$, and is stable.

The main idea of our approach is that both the strong and parity-violating
pion-nucleon couplings, g$_{\pi NN}$ and f$_{\pi NN}$, respectively, are
given by the lowest dimensional diagram and by a pion vacuum susceptibility,
$\chi_{\pi}$.
Therefore, by using the known value of g$_{\pi NN}$ one can determine
$\chi_{\pi}$ and thereby predict the value of f$_{\pi NN}$.
Using definitions for the susceptibilities given in Ref.\cite{HHK}, the 
QCD sum rule for the strong $\pi NN$ coupling is\cite{WYPH}

\begin{eqnarray}
& M_B^6 L^{-4/9}E_2 -{\chi_{\pi}} a  M_B^4 L^{2/9}E_1  
- {11\over 24}<g_c^2 G^2> M_B^2 E_0 +{4\over 3} a^2 L^{4/9}
+{1\over 3} m_0^2 a^2 {L^{-2/27}\over M_B^2} \nonumber\\
&= \{ g_{\pi NN} + B M_B^2\} \bar{\lambda}^2_N\;e^{-M^2/M^2_B}.
\end{eqnarray}
This equation is the same as Eq. (16) in Ref.\cite{HHK} except for changes
in coefficients of terms whose contributions are negligible within the
errors of the method, an addition of a dimension eight term, which also does
not alter our results, and an explicit inclusion of a monpole term from the
continuum. As was discussed in detail in Ref.\cite{HHK}, from the 
dominant first two terms on the left hand side of the above sum rule 
and by making use of the known value of g$_{\pi NN}$ we find the value
of  $\chi_\pi a$ is about -1.9 GeV$^2$; within the errors of both calculations,
this value is close to that
obtained in a three-point evaluation\cite{jk} using nonlocal condensates. 
It is notable that we have shown
explicitly that the large value of the strong $\pi NN$ coupling 
requires nonperturbative QCD.

In a subsequent sum rule calculation in which the chiral-quark-model value 
for the $\pi$-quark coupling constant, $g_{\pi q}$, and a value for 
$\chi_\pi a$ in the vicinity of $- 1.5\, GeV^2$ were used\cite{WYPH},
a value of $g_{\pi NN} = -(14.8 \pm 0.7)$ for $M_B = (1.10 \pm0.05)\, GeV$
was obtained. This is a fairly stable result with respect to the Borel mass.
An additional error of at least $\pm 2.0$ should be understood 
in view of the uncertainties involved in the various condensates.

We turn our attention to the weak p.v. pion-nucleon 
coupling, which is defined as follows \cite{DDH}
\begin{eqnarray}
{\cal L}_{\pi NN}^{p.v.} = {f_{\pi NN}\over \sqrt2} 
\bar\psi({\vec \tau}\times{\vec \pi})_3 \psi. 
\end{eqnarray}
Only charged pions can be emitted or absorbed. Here we modify the QCD 
sum rule which we obtained earlier\cite{HHK} (and which has some pathological 
behavior as the Borel mass increases) by multiplying both sides (LHS and
RHS) by
the factor $(p^2 - M^2)$ immediately before taking the Borel transform. This 
has helped to produce a QCD sum rule which is very well behaved.
\begin{eqnarray}
 &&{G_F\sin^2\theta_W({17\over 3} - \gamma)
\over 24 \pi^2}\; \bigl[ (4M^{10}_B - M^2 M_B^8) L^{-4/9}E_3 \nonumber\\
& -& 4 ( M_B^8 - {1\over 3} M^2 M_B^6)\chi_\pi\;aL^{-4/9} E_2 
 - (M_B^6 - {1\over 2} M^2 M_B^4) m_0^\pi\;a\;E_1 L^{-4/9}\bigr] \nonumber\\
& = & \{f_{\pi NN} +B' M_B^2\} \bar{\lambda}^2_N\; 2 M^2\;
      e^{-M^2/M^2_B}.
\end{eqnarray}
This sum rule still suffers from the fact that the contribution from
the gluon condensate diagrams is yet to be included; these diagrams are
much more complicated to evaluate although they are of the same order 
or smaller than uncertainties of our calculation. 

Numerically, we obtain $f_{\pi NN} = (3.04 \pm 0.01) \times 10^{-7}$ for
$M_B =(1.10 \pm 0.05)\, GeV$, a prediction which is even more stable than
the other sum rules (on the nucleon mass and the strong pion-nucleon
coupling). The uncertainties in the condensates and in the terms which have
been neglected amount to at least $\pm 0.5 \times 10^{-7}$. About 50 \% 
of the contribution to $f_{\pi NN}$
comes from the nonperturbative $\chi_\pi a$ term.  
The present prediction on $f_{\pi NN}$ is in much better agreement with that of
Desplanques, Donoghue and Holstein \cite{DDH} and also with that obtained
independently by one of us \cite{WYPH}.

In summary, our published Letter \cite{HHK} unfortunately contains
a sign error which completely changes the results. In particular, the sign in
front of the $(2/3) \chi_\pi a L^{-4/9} M_B^2 E_2 $ in Eq. (14) should be $-$ 
and not $+$. The first (lowest-dimensional) and second (non-perturbative) 
terms then add rather than subtract. 

    We have also evaluated two diagrams that we omitted in our earlier
work, namely Figs. 3d and 3e, which do appear in the external field method,
because it is an internal pion which couples to the gauge boson. However,
we find that these diagrams make a negligible contribution.

\vskip 0.5 true cm

\end{document}